%% file: main.tex
\documentclass[11pt]{article}
\usepackage{amsmath, amssymb, amsthm}
\usepackage[margin=1in]{geometry}
\usepackage{enumitem}
\usepackage{booktabs}
\usepackage{graphicx}
\usepackage{float}
\usepackage{needspace}
\usepackage[font=small]{caption}
\usepackage[authoryear,round]{natbib}
\usepackage{hyperref}
\usepackage{doi}

\theoremstyle{plain}
\newtheorem{thm}{Theorem}
\newtheorem{lem}{Lemma}
\newtheorem{cor}{Corollary}
\newtheorem{prop}{Proposition}

\theoremstyle{definition}
\newtheorem{defn}{Definition}
\newtheorem{assumption}{Assumption}
\newtheorem*{example}{Example}

\newcommand{\ind}{\mathbf{1}}
\newcommand{\AUC}{\mathrm{AUC}}
\newcommand{\BS}{\mathrm{BS}}
\newcommand{\BSS}{\mathrm{BSS}}
\newcommand{\AUCw}{\mathrm{AUC}_w}
\newcommand{\Fm}{F^{\mathrm{mid}}}

\title{One Curve, Different Weights: AUC, the Brier Score, Decision-Weighted AUC, and the Policy Youden Index}
\author{Yegor Baranovski\thanks{Email: \href{mailto:yegorbar01@gmail.com}{yegorbar01@gmail.com}. I thank Scott Nelson for posing the question that motivated this paper.}}
\date{\today}

\begin{document}
\maketitle

\begin{abstract}
AUC and the Brier score are standard summaries of a binary risk score, yet they can rank models differently. We
introduce a weighted AUC, $\mathrm{AUC}_\pi$, which weights each pair of observations by how differently a decision
maker with probability of acting $\pi$ treats them. We show that all three are built on weighted averages of one curve, the Youden index
$J(t) = \mathrm{TPR}(t) - \mathrm{FPR}(t)$ across cutoffs $t$, and that the averages differ in the weights they put on the cutoffs. Each
weights a cutoff by how much something rises across it: for AUC, the distribution of all scores; for the Brier score,
the true probability $\mathbb{E}[Y \mid S]$; for $\mathrm{AUC}_\pi$, the decision maker's probability of acting. Up to
normalization, $\mathrm{AUC}_\pi$ is the Brier skill score of the ideal forecast with its weight moved to the cutoffs the decision maker cares about. We also show that weighting pairs by
$|\pi(s) - \pi(s')|$ is, up to scale, the only pair weight fixed in advance that makes a pair-weighted AUC an average
of the Youden curve for every score distribution. As a ranking measure, AUC weights every model the same way.
\citet{hand2009measuring} shows that it can also be read as an average over cost thresholds, and in that reading each
model weights the thresholds differently. The numerator of $\mathrm{AUC}_\pi$, which we call the policy
Youden index, is the expected share of class 1 a fixed policy acts on minus the expected share of class 0. It judges every model by the
same policy and so avoids this; the full ratio $\mathrm{AUC}_\pi$ does not.
\end{abstract}

\input{01_introduction/introduction}
\input{02_setup/setup}
\input{03_theorems/theorems}
\input{04_characterization/characterization}
\input{05_coherence/coherence}
\input{08_conclusion/conclusion}

\bibliographystyle{plainnat}
\bibliography{refs}

\appendix
\input{09_appendix/appendix}

\end{document}

%% file: 01_introduction/introduction.tex
\section{Introduction}
\label{sec:introduction}

Models judged by AUC or the Brier score can be used to decide who to act on. A lender's default model decides who to
approve, a hospital's risk model who to treat, a bank's fraud model which
payments to block, and a weather service's rain forecast when to issue a warning. So the
natural question is what each measure rewards, in terms of the decisions the score supports.

This paper answers that question with one object, which we call the \textbf{Youden curve}. A cutoff $t$ acts on every observation
with a score above $t$. Its Youden index $J(t) = \mathrm{TPR}(t) - \mathrm{FPR}(t)$ is the share of class 1 it acts on
minus the share of class 0 \citep{youden1950index}. Each cutoff is a decision, and $J(t)$ measures its performance. $J$
does not depend on the base rate. A measure that does not commit to one cutoff must average $J$ across
cutoffs, and describe how much each cutoff counts. We show that three measures are built on exactly such averages, and that they
differ in the weights they put on the cutoffs. For a nondecreasing weight $w$ with values in $[0,1]$, let $T_w$ be a random cutoff with CDF $w$,
so $\mathbb{E}[J(T_w)]$ is the average of $J$ with each cutoff weighted by how much $w$ rises across it. Write $S_y$ for the score of a
class-$y$ observation, $r(s) = P(Y = 1 \mid S = s)$ for the true probability, $p$ for a forecast of it, $\pi(s)$ for the
decision maker's probability of acting on score $s$, and $\Fm$ for the CDF of the scores with ties split in half.
\begin{enumerate}[leftmargin=*]
\item \emph{AUC} weights each cutoff by how much the distribution of all scores rises across it. This is the same as
counting every share of observations acted on equally (Theorem~\ref{thm:auc}):
\[
\AUC = \tfrac12 + \mathbb{E}[J(T_{\Fm})].
\]
\item \emph{The Brier score} weights each cutoff by how much the true probability $r$ rises across it. This is the
same as counting every risk threshold equally, since the thresholds that lead to a cutoff span exactly that rise. It
adds a penalty for a forecast that is not ideal (Theorem~\ref{thm:brier}):
\[
\BS = \mathbb{E}[(p - r)^2] + \mathrm{Var}(Y)\bigl(1 - \mathbb{E}[J(T_r)]\bigr).
\]
For the ideal forecast $p = r$, the Brier skill score is $\mathbb{E}[J(T_r)]$ (Corollary~\ref{cor:bss}).
\item \emph{$\mathrm{AUC}_\pi$} weights each pair of a class-0 and a class-1 observation, with scores $s$ and $s'$, by
$|\pi(s) - \pi(s')|$, how differently the decision maker treats them. It weights each cutoff by how much $\pi$ rises across it
(Theorem~\ref{thm:aucw}):
\[
\mathrm{AUC}_\pi = \frac12 + \frac{\mathbb{E}[J(T_\pi)]}{2\,\mathbb{E}|\pi(S_1) - \pi(S_0)|}.
\]
\end{enumerate}
All three follow from one identity, $\mathbb{E}[J(T_w)] = \mathbb{E}[w(S_1)] - \mathbb{E}[w(S_0)]$
(Lemma~\ref{lem:cutoff}). All results hold for discrete scores with ties.

Theorem~\ref{thm:char} explains this choice of pair weight. If a pair-weighted AUC with weights fixed in advance is an
average of the Youden curve for every score distribution, its pair weight must be a multiple of $|w(s) - w(s')|$ for
some $w$, which then dictates its cutoff weight. So choosing pair weights is choosing cutoff weights. Plain AUC gives
every pair equal weight, and its cutoff weights come from the model's own scores, so, in terms of risk thresholds,
they change when the model changes.

Since each measure averages $J$ over cutoffs, choosing a measure means choosing how much each decision matters. AUC
weights equally acting on the top 10\%, the top 20\%, and so on. The Brier score weights equally acting on everyone
whose true probability is above 10\%, above 20\%, and so on. It also charges for a forecast that is not ideal, even when the decisions depend
only on the ranking. $\mathrm{AUC}_\pi$ gives each decision the weight the decision maker gives it, and like AUC it
does not need the score to be a probability. Up to
normalization, $\mathrm{AUC}_\pi$ is the Brier skill score of the ideal forecast with its weight moved from where $r$
rises to where $\pi$ rises; when $\pi$ is linear in $r$, it is exactly a rescaled Brier skill score
(Proposition~\ref{prop:link}). The normalization does depend on the model, so to compare models we use the numerator, which
we call the policy Youden index, $J_\pi = \mathbb{E}[J(T_\pi)]$: the expected share of class 1 the policy acts on minus the
expected share of class 0 (Section~\ref{sec:coherence}).

Section~\ref{sec:setup} sets up the Youden curve. Section~\ref{sec:theorems} states the cutoff lemma and the three
results. Section~\ref{sec:why} characterizes the pair weights. Section~\ref{sec:coherence} takes up Hand's coherence
critique with a worked example, shows why models should be compared by the policy Youden index, and
extends the point to comparisons across groups. Section~\ref{sec:conclusion} concludes, and the proofs are in
Appendix~\ref{sec:proofs}.

\subsection{Related work}

\paragraph{Measures as averages over cutoffs.} The closest work is \citet{hernandezorallo2012unified}, who write AUC, the
Brier score and other measures as expected classification loss under different rules for choosing the cutoff
\citep[for AUC, see also][]{flach2011coherent}. \citet{hand2009measuring} shows that AUC is a linear function of the
average minimum loss over a cost distribution that depends on the classifier itself, and proposes the H-measure with a
fixed cost distribution instead. \citet{schervish1989general} shows that every proper scoring rule, the Brier score
among them, is a mixture of cost-weighted losses over cutoffs on the probability scale.
We average a different quantity, $J$, which does not depend on the base rate or on costs, so every measure becomes an
average of the same curve. Theorem~\ref{thm:auc} then gives Hand's diagnosis in Youden terms: AUC's weight is the
model's own score distribution, which puts equal weight on every share acted on, and Theorem~\ref{thm:brier} keeps Schervish's uniform thresholds but applies them to the true
probability $r$ rather than the forecast.

\paragraph{Earlier averages of $J$.}
Averages of $J$ have appeared for two weights. The first is uniform over risk thresholds, the Brier weight of
Corollary~\ref{cor:bss}: \citet{pencina2008evaluating} describe the integrated discrimination improvement, a difference
of two discrimination slopes, as an integrated difference in Youden indices. The second is the pooled score distribution,
where AUC is $\tfrac12$ plus the area under the KS curve \citep{adeodato2022kolmogorov}. What
is new here is the single identity for an arbitrary weight, which recovers both as special cases. We also introduce
$\mathrm{AUC}_\pi$ as a measure of discrimination at the decision maker's own cutoffs, and characterize its pair
weights. Its numerator, the policy Youden index $J_\pi$, judges every model by the same policy, so it avoids Hand's
objection without a cost model or a probability forecast, while the ratio $\mathrm{AUC}_\pi$ does not.

\paragraph{Decision-weighted measures.} Other measures weight a payoff by a distribution of decision thresholds. One
such payoff is the net benefit of \citet{vickers2006decision}. \citet{talluri2016weighted} weight net benefit by a
distribution of thresholds estimated from observed clinical decisions, and \citet{zhu2025weighted} weight the Brier
score by a distribution of thresholds. These measures also depend on calibration, not only on the ranking.
Weighted AUCs already exist. \citet{li2010weighted} weight the ROC curve by a density on specificity, and
\citet{wieand1989family} average sensitivity over a distribution on that same axis. Those weights sit on the
false-positive axis. $\mathrm{AUC}_\pi$ weights each pair by how differently the policy treats its two scores,
which puts the weight on the score scale, so the score need not be a probability and only discrimination is
measured. Theorem~\ref{thm:char} says this is the only pair weight, fixed in advance, that makes the result an
average of $J$.

%% file: 02_setup/setup.tex
\section{Setup}
\label{sec:setup}

Each observation has a score $S$ and a binary label $Y \in \{0,1\}$, with $\rho = P(Y = 1) \in (0,1)$. The score is
the output of the model being evaluated, with higher scores
meaning class 1 is more likely. It need not be a probability. Comparing models means comparing different scores $S$
for the same observations. A decision maker
either takes an \textbf{action} or does not; the action is aimed at class 1. Write $F_y(t) = P(S \le t \mid Y = y)$ for
the CDF of class $y$, and $F(t) = P(S \le t) = (1 - \rho)F_0(t) + \rho F_1(t)$ for the CDF of all scores. The score
may have any distribution; in particular it may be discrete, so ties are allowed. Write
$\Fm(t) = P(S < t) + \tfrac12 P(S = t)$ for the \textbf{mid-point} version of $F$, the share of scores strictly below $t$
plus half the share tied at $t$, and define $\Fm_y$ the same way within class $y$. For continuous scores $\Fm = F$.
Let $S_0 \sim F_0$ and $S_1 \sim F_1$ be independent: the scores of a random class-0 and a random class-1 observation.

\begin{assumption}[Monotonicity]\label{ass:monotone}
$r(s) = P(Y = 1 \mid S = s)$ is nondecreasing in $s$.
\end{assumption}

Monotonicity is a condition on the two classes together, not on $F_0$ or $F_1$ alone. With class densities $f_0$ and
$f_1$ it is equivalent, by Bayes' rule, to the likelihood ratio $f_1/f_0$ being nondecreasing: the higher the score, the
more it looks like class 1. It is what makes acting on high scores sensible.

\subsection{The ROC curve and the Youden curve}

\begin{figure}[H]
\centering
\includegraphics[width=\textwidth]{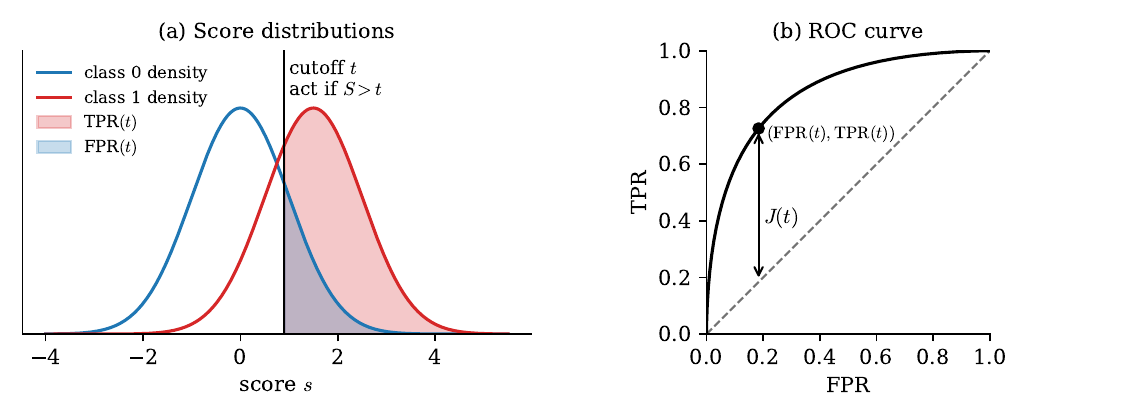}
\caption{(a) Class-0 scores $\sim N(0,1)$ and class-1 scores $\sim N(1.5,1)$. At cutoff $t$, the action reaches the
shaded share of each class: $\mathrm{TPR}(t)$ for class 1 and $\mathrm{FPR}(t)$ for class 0. (b) Moving $t$ from
$+\infty$ to $-\infty$ traces the ROC curve. The Youden index $J(t)$ is the vertical gap between the curve and the
diagonal.}
\label{fig:roc}
\end{figure}

\begin{figure}[H]
\centering
\includegraphics[width=\textwidth]{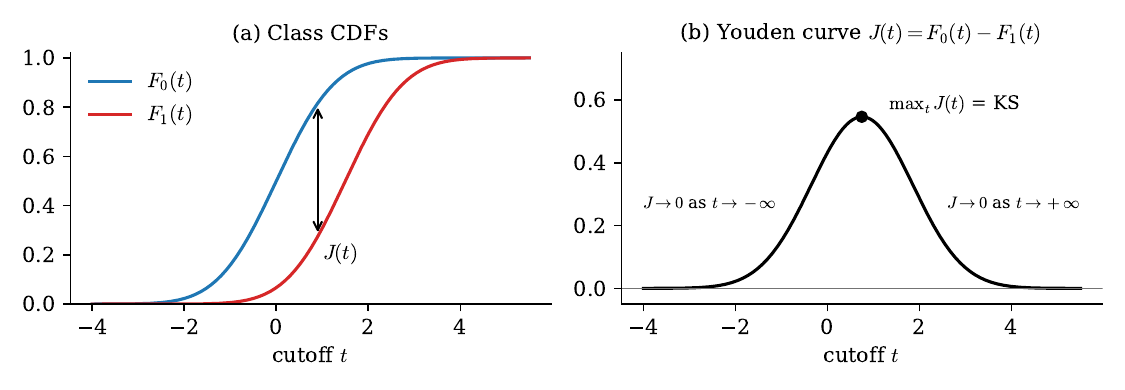}
\caption{Same scores as Figure~\ref{fig:roc}. (a) $J(t)$ is the vertical gap between the class CDFs. (b) The Youden
curve is zero at both extremes and peaks at the best single cutoff; its maximum is the Kolmogorov--Smirnov (KS)
statistic.}
\label{fig:youden}
\end{figure}

A score is used through cutoffs: the decision maker acts on every observation whose score is high enough. The cutoff
``act when $S > t$'' acts on a share of each class, the \textbf{true positive rate} and the \textbf{false positive
rate}:
\[
\mathrm{TPR}(t) = P(S > t \mid Y = 1) = 1 - F_1(t), \qquad \mathrm{FPR}(t) = P(S > t \mid Y = 0) = 1 - F_0(t).
\]
As $t$ moves from $+\infty$ to $-\infty$, the point $(\mathrm{FPR}(t), \mathrm{TPR}(t))$ traces the
\textbf{ROC curve} from $(0,0)$ to $(1,1)$ (Figure~\ref{fig:roc}). The \textbf{Youden index} of the cutoff
is the share of class 1 it acts on minus the share of class 0,
\[
J(t) = \mathrm{TPR}(t) - \mathrm{FPR}(t) = \bigl(1 - F_1(t)\bigr) - \bigl(1 - F_0(t)\bigr) = F_0(t) - F_1(t),
\]
the vertical gap between the ROC curve and the diagonal. The diagonal is the line $\mathrm{TPR} = \mathrm{FPR}$, so
directly below the point $(\mathrm{FPR}(t), \mathrm{TPR}(t))$ it is at height $\mathrm{FPR}(t)$. As a function of $t$, we call $J$ the \textbf{Youden curve}
(Figure~\ref{fig:youden}). It is zero at both extremes, $J(\pm\infty) = 0$, since acting on everyone or on no one does
not separate the classes. $J(t)$ is the gap between the two class CDFs at cutoff $t$. Since $J \ge 0$ under
Assumption~\ref{ass:monotone} (Lemma~\ref{lem:Jnonneg}), the largest $J(t)$ is the two-sample Kolmogorov--Smirnov (KS)
statistic \citep{kolmogorov1933sulla,smirnov1939estimation}.

\begin{lem}[$J$ is nonnegative]\label{lem:Jnonneg}
Under Assumption~\ref{ass:monotone}, $F_1(t) \le F(t) \le F_0(t)$ for every $t$. Hence $J(t) \ge 0$ at every cutoff.
\end{lem}

The proof is in \hyperref[prf:Jnonneg]{Appendix~\ref*{sec:proofs}}. A higher score never makes class 1 less likely, so
the scores at or below any cutoff hold a smaller share of class 1 than of class 0, and the ROC curve never falls below
the diagonal. Three facts make $J$ the natural measure of how well one cutoff separates the classes.
\begin{itemize}[leftmargin=*]
\item \emph{It is measured against random acting.} Acting on each observation with the same probability, ignoring the
score, reaches the same share of each class, so $J = 0$. So $J$ is the extra share of class 1 a cutoff reaches compared
with random acting that reaches the same share of class 0.
\item \emph{It uses only shares within each class,} so it does not depend on the base rate $\rho$.
\item \emph{It counts pairs.} Pair a random class-0 observation with a random class-1 observation, so that
$P(S_0 \le t) = F_0(t)$ and $P(S_1 \le t) = F_1(t)$. Splitting each by where the other member falls,
\[
F_0(t) = P(S_0 \le t,\, S_1 \le t) + P(S_0 \le t < S_1), \qquad
F_1(t) = P(S_0 \le t,\, S_1 \le t) + P(S_1 \le t < S_0).
\]
Subtracting, the first terms cancel:
\[
J(t) = P(S_0 \le t < S_1) - P(S_1 \le t < S_0),
\]
the share of pairs that the cutoff splits correctly (class 1 acted on, class 0 not) minus the share it splits
incorrectly. AUC also counts pairs, which is why the two are linked.
\end{itemize}

%% file: 03_theorems/theorems.tex
\section{Main results}
\label{sec:theorems}

Writing $\mathbb{E}[J(T_w)]$ for the weighted average of $J$ over the cutoffs of a weight $w$
(Section~\ref{sec:tool}), the main results are:
\begin{center}
\begin{tabular}{@{}llll@{}}
\toprule
Measure & Identity & Weight $w$ & Cutoff weighted by \\
\midrule
AUC & $\AUC = \tfrac12 + \mathbb{E}[J(T_{\Fm})]$ & distribution of scores $\Fm$ & how much $\Fm$ rises \\[2pt]
Brier & $\BSS = \mathbb{E}[J(T_r)]$ (ideal $p = r$) & true probability $r$ & how much $r$ rises \\[2pt]
$\mathrm{AUC}_\pi$ & $\mathrm{AUC}_\pi = \tfrac12 + \dfrac{\mathbb{E}[J(T_\pi)]}{2\,\mathbb{E}|\Delta_\pi|}$ & decision maker's weight $\pi$ & how much $\pi$ rises \\
\bottomrule
\end{tabular}
\end{center}
Section~\ref{sec:tool} turns a weight into an average of $J$, and Sections~\ref{sec:auc}--\ref{sec:link} derive the
three measures and how $\mathrm{AUC}_\pi$ relates to the Brier score.

\subsection{\texorpdfstring{Averaging $J$ over a weight's cutoffs}{Averaging J over a weight's cutoffs}}\label{sec:tool}

\begin{defn}[Weight]\label{def:weight}
A \textbf{weight} is a nondecreasing, right-continuous function $w : \mathbb{R} \to [0,1]$. We interpret $w(s)$ as the
probability of acting on an observation with score $s$.
\end{defn}

A weight decides how much each cutoff counts in the average. Nondecreasing means higher scores are acted on at least
as often. A smooth $w$ is already right-continuous.

\needspace{6\baselineskip}
\begin{lem}[Weights on cutoffs]\label{lem:weights}\leavevmode
\begin{enumerate}[label=(\alph*), leftmargin=*]
\item \emph{Smooth case.} If $w$ has a density $w'$, then
\[
\mathbb{E}[w(S_1)] - \mathbb{E}[w(S_0)] = \int J(t)\, w'(t)\, dt.
\]
\item \emph{Discrete case.} If the scores take finitely many values $s_1 < \dots < s_n$, then
\[
\mathbb{E}[w(S_1)] - \mathbb{E}[w(S_0)] = \sum_{k=2}^{n} \bigl[w(s_k) - w(s_{k-1})\bigr]\, J(s_{k-1}).
\]
\end{enumerate}
\end{lem}

The proof is in \hyperref[prf:weights]{Appendix~\ref*{sec:proofs}}. The left side is how much more often $w$ acts on
class 1 than on class 0. The right side is the Youden curve weighted by $w'$: cutoff $t$ counts in proportion to how
fast $w$ rises there, and cutoffs where $w$ is flat don't count at all. The discrete case is the same with $w'(t)\,dt$
replaced by the rise $w(s_k) - w(s_{k-1})$, and the integral replaced by a sum.

\begin{defn}[Random cutoff]\label{def:cutoff}
Define a random variable $T_w$ with values in $[-\infty, \infty]$, interpreted as a random cutoff, independent of
$(S_0,S_1)$, with CDF
\[
F_{T_w}(t) = P(T_w \le t) =
\begin{cases}
w(-\infty) & t = -\infty,\\
w(t) & t \in \mathbb{R},\\
1 & t = +\infty,
\end{cases}
\]
where $w(\pm\infty)$ are the limits of $w$. The cutoff $-\infty$ acts on everyone and $+\infty$ on no one, so
$J(\pm\infty) = 0$.
\end{defn}

\begin{figure}[H]
\centering
\includegraphics[width=\textwidth]{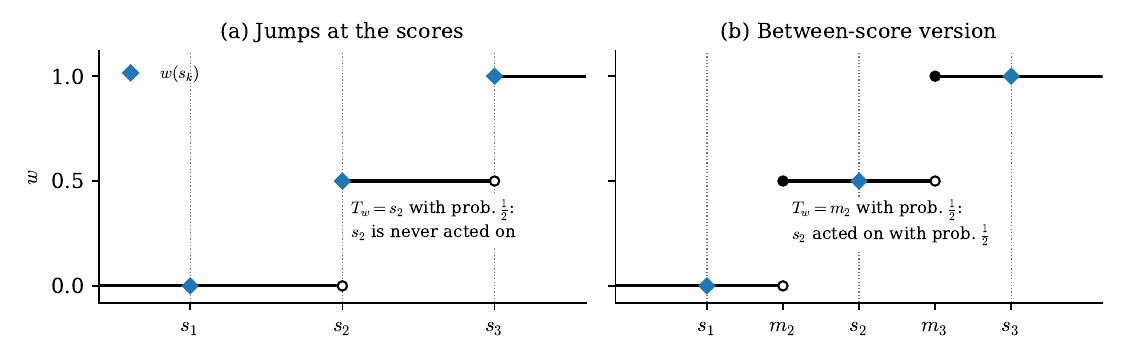}
\caption{A weight with $w(s_1) = 0$, $w(s_2) = \tfrac12$, $w(s_3) = 1$ (diamonds). Filled circles are values of $w$ and
open circles are limits from the left. (a) If $w$ jumps at the scores, the random cutoff $T_w$ lands on $s_2$ or $s_3$,
each with probability $\tfrac12$, and ``act when $S > T_w$'' never acts on $s_2$, although $w(s_2) = \tfrac12$. (b) Moving
each jump to the midpoint keeps the values at the scores, and the rule acts on each score $s_k$ with probability
$w(s_k)$.}
\label{fig:between}
\end{figure}

$T_w$ is a cutoff drawn at random with CDF $w$, and the rule acts when $S > T_w$. With discrete scores
$s_1 < \dots < s_n$, call a weight \textbf{between-score} if it has no jump at any $s_k$. If $w$ jumps at a score,
$T_w$ can land on that score, and the rule then does not act there (Figure~\ref{fig:between}a). We deal with this by
moving each jump to the midpoint between neighbouring scores, which keeps the values at the scores and removes the
problem (Figure~\ref{fig:between}b). We only care about $w$ at the actual scores, so this is harmless. With discrete
scores we always use this between-score version. The problem comes from the strict rule. Under ``act when
$S \ge T_w$'' it does not arise, but then $\mathrm{TPR}(t) = 1 - P(S_1 < t)$ and $J$ is no longer $F_0 - F_1$. We keep
``act when $S > t$'' so that both keep their standard form.

For a between-score or continuous weight, $P(T_w < s) = w(s)$ at every score $s$, so acting when $s > T_w$ is the same
as acting on score $s$ with probability $w(s)$. Random cutoffs are standard: the threshold
choice methods of \citet{hernandezorallo2012unified} include cutoffs drawn uniformly over the range of scores and
over predicted positive rates. What we add is the general weight $w$ and the identity below.

\begin{lem}[Cutoff lemma]\label{lem:cutoff}
In either case of Lemma~\ref{lem:weights}, with $w$ between-score in the discrete case,
$\;\mathbb{E}[J(T_w)] = \mathbb{E}[w(S_1)] - \mathbb{E}[w(S_0)]$.
\end{lem}

The proof is in \hyperref[prf:cutoff]{Appendix~\ref*{sec:proofs}}. The left side is the weighted average of $J$ over
the cutoffs of $w$, and the right side is how much more often $w$ acts on class 1 than on class 0. By
Lemma~\ref{lem:weights}, both equal
\[
\mathbb{E}[J(T_w)] =
\begin{cases}
\displaystyle\int J(t)\, w'(t)\, dt & \text{if $w$ is smooth},\\[10pt]
\displaystyle\sum_{k=2}^{n} \bigl[w(s_k) - w(s_{k-1})\bigr]\, J(s_{k-1}) & \text{if the scores are discrete}.
\end{cases}
\]

For one weight the lemma is known. When $w$ is a predicted risk, $\mathbb{E}[w(S_1)] - \mathbb{E}[w(S_0)]$ is the
average predicted risk in class 1 minus that in class 0, called the discrimination slope. \citet{pencina2008evaluating}
show that the slope is the integral of $J$ over risk thresholds in $(0,1)$, so the difference between two models'
slopes is an integrated difference of their Youden indices. Lemma~\ref{lem:cutoff} states it for every weight, and
every result below is one choice of $w$.

The results below hold in both cases of Lemma~\ref{lem:weights}. With discrete scores, $\Fm$, $r$ and $\pi$ enter only
through their values at the scores, and we use their between-score versions.

\subsection{AUC}\label{sec:auc}

\begin{defn}[AUC]\label{def:auc}
$\AUC = P(S_0 < S_1) + \tfrac12 P(S_0 = S_1)$: the probability that a random class-1 observation outscores a random
class-0 observation, with ties counted as half. This equals the area under the ROC curve, with its points joined by
straight lines if the scores are discrete \citep{bamber1975area,hanley1982meaning}.
\end{defn}

Take the weight $w = \Fm$. With scores that have a density $f$, $T_{\Fm}$ is a cutoff distributed like the mix of
scores, so a cutoff region counts in proportion to how many observations sit there:
\[
\mathbb{E}[J(T_{\Fm})] = \int J(t)\, f(t)\, dt.
\]

\begin{thm}[AUC]\label{thm:auc}
$\AUC = \tfrac12 + \mathbb{E}[J(T_{\Fm})]$.
\end{thm}

The proof is in \hyperref[prf:auc]{Appendix~\ref*{sec:proofs}}. AUC is the average Youden index with each cutoff weighted
by how much the distribution of all scores rises across it. The mix of the two classes does not matter, as the same holds for
$w = (1 - \lambda)\Fm_0 + \lambda \Fm_1$ with any $\lambda \in [0,1]$, and $\Fm$ is $\lambda = \rho$. For continuous
scores $F(T_F)$ is uniform on $(0,1)$, so the share of observations acted on, $1 - F(T_F)$, is uniform too. Therefore,
Theorem~\ref{thm:auc} says that $\AUC - \tfrac12$ is the average $J$ over decision makers who act on a uniformly random
share of the population. The continuous case is known, stated as expected loss
\citep{flach2011coherent,hernandezorallo2012unified}. The closest is that AUC is $\tfrac12$ plus the area under the KS
curve \citep{adeodato2022kolmogorov}. That curve is the
Youden curve plotted against the pooled share $F(t)$ instead of $t$. Equal widths on that axis are equal shares of observations, so its area is the average of
$J$ that Theorem~\ref{thm:auc} gives.

\subsection{The Brier score}\label{sec:brier}

\begin{defn}[Brier score]\label{def:brier}
Let $p = p(S) \in [0,1]$ be a forecast of $P(Y = 1 \mid S)$. The \textbf{Brier score} is
$\BS = \mathbb{E}\bigl[(p - Y)^2\bigr]$, and the forecast is \textbf{ideal} if $p = r = \mathbb{E}[Y \mid S]$. The
\textbf{Brier skill score} is $\BSS = 1 - \BS / \mathrm{Var}(Y)$.
\end{defn}

The ideal forecast minimizes $\BS$ among all forecasts based on $S$. By the decomposition of \citet{murphy1973new},
grouping observations by their score $S$ rather than by their forecast $p$,
\[
\BS = \mathbb{E}\bigl[(p - r)^2\bigr] - \mathrm{Var}(r) + \mathrm{Var}(Y).
\]
For the ideal forecast, $\BS = \mathrm{Var}(Y) - \mathrm{Var}(r)$ and $\BSS = \mathrm{Var}(r)/\mathrm{Var}(Y)$. Under
Assumption~\ref{ass:monotone}, $r$ is itself a weight.

\begin{thm}[Brier score]\label{thm:brier}
Under Assumption~\ref{ass:monotone},
\[
\BS = \mathbb{E}\bigl[(p - r)^2\bigr] + \mathrm{Var}(Y)\bigl(1 - \mathbb{E}[J(T_r)]\bigr).
\]
\end{thm}

\begin{cor}\label{cor:bss}
For the ideal forecast, $\BSS = \mathbb{E}[J(T_r)]$.
\end{cor}

The proofs are in \hyperref[prf:brier]{Appendix~\ref*{sec:proofs}}. The Brier score weights each cutoff by how much the
true probability $r$ rises across it, and adds a separate term $\mathbb{E}[(p - r)^2]$ for how far the forecast is from
ideal.

On the score scale, the axis of cutoffs $t$, the weights come from $r$. On the probability scale, the axis of risk
thresholds $u = r(t)$, they are equal. To see this, picture decision
makers who each act on everyone whose true probability is above their own risk threshold $u$. Many thresholds lead
to the same decision. Take three scores with $r(s_1) = 0.1$, $r(s_2) = 0.3$ and $r(s_3) = 0.9$. Every $u$ between 0.1
and 0.3 acts on the top two scores, which is cutoff $s_1$. Every $u$ between 0.3 and 0.9 acts on the top score only,
which is cutoff $s_2$. A $u$ below 0.1 acts on everyone and a $u$ above 0.9 on no one, and both give $J = 0$. If every
threshold $u$ in $[0,1]$ counts equally, cutoff $s_1$ gets weight $0.3 - 0.1 = 0.2$ and cutoff $s_2$ gets
$0.9 - 0.3 = 0.6$. These are the rises in $r$, the weights of Corollary~\ref{cor:bss}. So the Brier skill score of the
ideal forecast is the average Youden index over risk thresholds spread evenly on $[0,1]$. A threshold of 90\% counts as
much as one of 5\%, even for a decision maker who only uses one of them. $\mathrm{AUC}_\pi$ instead puts the weight where the decision
maker acts.

The Brier skill score of the ideal forecast,
$\mathrm{Var}(r)/\mathrm{Var}(Y)$, equals the gap in average true probability between the two classes,
$\mathbb{E}[r \mid Y = 1] - \mathbb{E}[r \mid Y = 0]$. The proof of Theorem~\ref{thm:brier} in
\hyperref[prf:brier]{Appendix~\ref*{sec:proofs}} shows this directly. Corollary~\ref{cor:bss} puts it on the score
scale, and the comparison needs this form. There it is an average of the same Youden curve $J(t)$ as AUC and
$\mathrm{AUC}_\pi$, so the three measures can be compared cutoff by cutoff and differ only in their weights.

This is close to the decision reading of \citet{schervish1989general}, which also averages over uniform thresholds.
Schervish thresholds the forecast $p$ and averages a cost-weighted loss; Theorem~\ref{thm:brier} thresholds the true
probability $r$, averages the unweighted $J$, and puts the forecast's error in the separate term
$\mathbb{E}[(p - r)^2]$.

\needspace{10\baselineskip}
\subsection{\texorpdfstring{$\AUCw$}{AUC w}}\label{sec:aucw}

\begin{defn}[$\AUCw$]\label{def:aucw}
AUC is the share of correctly ordered pairs, with every pair weighted equally:
\[
\AUC = \mathbb{E}\bigl[\ind\{S_0 < S_1\} + \tfrac12 \ind\{S_0 = S_1\}\bigr].
\]
For a weight $w$, let $\Delta_w = w(S_1) - w(S_0)$, how differently $w$ treats the two members of a pair. Weighting each
pair by $|\Delta_w|$ instead of $1$, and dividing by the total weight so the result is still a share, gives, assuming
$\mathbb{E}|\Delta_w| > 0$,
\[
\AUCw = \frac{\mathbb{E}\bigl[|\Delta_w|\, \ind\{S_0 < S_1\}\bigr]}{\mathbb{E}|\Delta_w|}.
\]
Tied pairs have $\Delta_w = 0$, so they need no tie term.
\end{defn}

The measure of interest is $\mathrm{AUC}_\pi$, where $\pi(s)$ is the probability that the decision maker acts on an
observation with score $s$. We take $\pi$ to be nondecreasing, so it is a weight. A pair the decision maker treats the
same way, both acted on or both not, cannot be misranked in any way that matters to that decision maker, and gets no
weight.

\begin{thm}[$\AUCw$]\label{thm:aucw}
For any weight $w$ with $\mathbb{E}|\Delta_w| > 0$,
\[
\AUCw = \frac12 + \frac{\mathbb{E}[J(T_w)]}{2\,\mathbb{E}|\Delta_w|}.
\]
\end{thm}

The proof is in \hyperref[prf:aucw]{Appendix~\ref*{sec:proofs}}. $\mathrm{AUC}_\pi$ therefore weights each cutoff by how
much the decision maker's probability of acting rises across it. Under Assumption~\ref{ass:monotone}, $J \ge 0$ at every
cutoff (Lemma~\ref{lem:Jnonneg}), so AUC and $\AUCw$ are at least $\tfrac12$ and the $\BSS$ of the ideal forecast is at
least $0$.

For the decision maker's policy we call the numerator the \textbf{policy Youden index},
\[
J_\pi = \mathbb{E}[J(T_\pi)] = \mathbb{E}[\pi(S_1)] - \mathbb{E}[\pi(S_0)],
\]
the expected share of class 1 the policy acts on minus the expected share of class 0. For a policy that acts above one cutoff
$t$, it is $J(t)$. $J_\pi$ is the Youden index of the randomized policy that acts on score $s$ with probability $\pi(s)$. It
needs no costs and no forecast probability; a decision maker with a known cost ratio $k$ can use
$\mathbb{E}[\mathrm{TPR}(T_\pi) - k\,\mathrm{FPR}(T_\pi)] = \mathbb{E}[\pi(S_1)] - k\,\mathbb{E}[\pi(S_0)]$ instead.
Section~\ref{sec:coherence} shows that models should be compared by $J_\pi$ rather than by
$\mathrm{AUC}_\pi$.

\subsection{Link to the Brier score}\label{sec:link}

\begin{prop}[Link]\label{prop:link}
$\AUCw$ is unchanged if $w$ is replaced by $\alpha + \beta w$ with $\beta > 0$. Hence, under
Assumption~\ref{ass:monotone}, if $\pi = \alpha + \beta r$ with $\beta > 0$,
\[
\mathrm{AUC}_\pi = \mathrm{AUC}_r = \frac12 + \frac{\mathbb{E}[J(T_r)]}{2\,\mathbb{E}|r(S_1) - r(S_0)|},
\]
and the numerator is the Brier skill score of the ideal forecast.
\end{prop}

The proof is in \hyperref[prf:link]{Appendix~\ref*{sec:proofs}}. So when the decision maker's probability of acting is
linear in the true probability, $\mathrm{AUC}_\pi$ is the Brier skill score of the ideal forecast, rescaled. When it is
not, $\mathrm{AUC}_\pi$ moves the weight to the cutoffs where decisions change.

%% file: 04_characterization/characterization.tex
\section{\texorpdfstring{Why weight pairs by $|w(s) - w(s')|$}{Why weight pairs by |w(s) - w(s')|}}
\label{sec:why}

One could weight pairs by some other function of $w$. To compare choices, define a general pair-weighted AUC.

\begin{defn}[Pair-weighted AUC]\label{def:aucW}
A \textbf{pair weight} is a bounded symmetric function $W(s, s') \ge 0$. Assuming $\mathbb{E}[W(S_0, S_1)] > 0$,
\[
\mathrm{AUC}_W = \frac{\mathbb{E}\bigl[W(S_0, S_1)\bigl(\ind\{S_0 < S_1\} + \tfrac12 \ind\{S_0 = S_1\}\bigr)\bigr]}{\mathbb{E}[W(S_0, S_1)]}.
\]
\end{defn}

$W \equiv 1$ gives AUC, and $W(s, s') = |w(s) - w(s')|$ gives $\AUCw$.

\begin{lem}[Pair form]\label{lem:pair}
For any pair weight $W$,
\[
\mathrm{AUC}_W = \frac12 + \frac{\mathbb{E}\bigl[W(S_0, S_1)\bigl(\ind\{S_0 < S_1\} - \ind\{S_1 < S_0\}\bigr)\bigr]}{2\,\mathbb{E}[W(S_0, S_1)]}.
\]
\end{lem}

The proof is in \hyperref[prf:pair]{Appendix~\ref*{sec:proofs}}. The numerator is the weight of correctly ranked pairs
minus the weight of incorrectly ranked pairs.

Averages of $J$ are pair-weighted in exactly this form. Recall from Section~\ref{sec:setup} that $J$ counts pairs,
$J(t) = P(S_0 \le t < S_1) - P(S_1 \le t < S_0)$. Replacing $t$ by $T_w$, which is independent of the pair,
\[
\mathbb{E}[J(T_w)] = P(S_0 \le T_w < S_1) - P(S_1 \le T_w < S_0).
\]
Fix a pair with class-0 score $s$ and class-1 score $s'$. If $s < s'$, $T_w$ splits it correctly when it lands in
$[s, s')$, which has probability $P(s \le T_w < s') = w(s') - w(s)$. If $s > s'$, it splits it incorrectly with
probability $w(s) - w(s')$, and if $s = s'$ it never splits it. Averaging over pairs,
\[
\mathbb{E}[J(T_w)] = \mathbb{E}\bigl[|w(S_1) - w(S_0)|\bigl(\ind\{S_0 < S_1\} - \ind\{S_1 < S_0\}\bigr)\bigr],
\]
the numerator of Lemma~\ref{lem:pair} with $W(s, s') = |w(s) - w(s')|$. So averaging $J$ over $T_w$ weights each pair
by how much of $T_w$ lies between its two scores. Theorem~\ref{thm:char} shows this is the only pair weight that
arises this way.

\begin{thm}[Characterization]\label{thm:char}
Let $W$ be a pair weight, $w: \mathbb{R} \to [0,1]$ nondecreasing and $c > 0$, all fixed in advance. Then
\[
\mathbb{E}\bigl[W(S_0, S_1)\bigl(\ind\{S_0 < S_1\} - \ind\{S_1 < S_0\}\bigr)\bigr] = c\,\mathbb{E}[J(T_w)]
\]
for all distributions $F_0, F_1$ covered by Lemma~\ref{lem:weights} (all of them if $w$ is smooth, all discrete ones
otherwise) if and only if $W(s, s') = c\,|w(s) - w(s')|$ for all $s \ne s'$.
\end{thm}

The proof is in \hyperref[prf:char]{Appendix~\ref*{sec:proofs}}. By Lemma~\ref{lem:pair} the left side is the numerator
of $\mathrm{AUC}_W$, so a pair-weighted AUC is an average of the Youden curve exactly when it is $\AUCw$. The weights
must not depend on $F_0$ and $F_1$. Plain AUC ($W \equiv 1$) fails this test: it is an average of $J$ in
Theorem~\ref{thm:auc} only because its weight there, $\Fm$, is built from the score distribution itself.

%% file: 05_coherence/coherence.tex
\section{Coherence across models and groups}
\label{sec:coherence}

\citet{hand2009measuring} objects that AUC measures different models with different rulers. Theorem~\ref{thm:auc}
shows why. A lender's costs set a risk threshold: approve everyone whose chance of repaying is above some level $u$.
By Theorem~\ref{thm:auc}, AUC weights cutoffs by where the observations are, so on the risk scale it weights each
threshold $u$ by how many observations have true probability near $u$. That distribution differs between models: a
threshold of 90\% may approve the top 60\% of applicants under one model and the top 40\% under another. So the weight
AUC puts on a given cost differs between models. By Corollary~\ref{cor:bss}, the Brier skill score of the ideal
forecast does not have this problem, since it weights every risk threshold equally under every model. \citet{flach2011coherent} reply that AUC is linear in expected loss when the cutoff is set at the score of a randomly
chosen observation, rather than at the best cutoff for the costs, and the costs are drawn uniformly. Neither
rule depends on the model, so as a ranking measure AUC weights every model the same way. Read as an average over cost
thresholds, it does not as each model puts its own weights on the thresholds, so comparing two AUCs compares averages
taken with different weights. Hand's H-measure fixes the weights on costs in advance instead.

With $n$ distinct scores, write $J_k$ for the Youden index of acting on the top $k$ observations. Then
Theorem~\ref{thm:auc} becomes
\begin{equation}\label{eq:share}
\AUC = \frac12 + \frac1n\,(J_1 + \dots + J_{n-1}),
\end{equation}
the average $J$ over acting on $1, 2, \dots, n-1$ of the $n$ observations (derivation in
\hyperref[prf:share]{Appendix~\ref*{sec:proofs}}).

\begin{example}
A lender scores ten applicants, five who repay ($Y = 1$) and five who default ($Y = 0$), with two models. From the
highest score to the lowest, the labels are
\[
\text{Model A: } 1\,0\,1\,1\,1\,0\,1\,0\,0\,0, \qquad \text{Model B: } 1\,1\,1\,0\,0\,0\,1\,0\,1\,0.
\]
Approving the top $k$ gives $J_k = (\text{repayers approved})/5 - (\text{defaulters approved})/5$:
\begin{center}
\begin{tabular}{@{}lccccccccc@{}}
\toprule
$k$ & 1 & 2 & 3 & 4 & 5 & 6 & 7 & 8 & 9 \\
\midrule
Model A & .2 & 0 & .2 & .4 & .6 & .4 & .6 & .4 & .2 \\
Model B & .2 & .4 & .6 & .4 & .2 & 0 & .2 & 0 & .2 \\
\bottomrule
\end{tabular}
\end{center}
By the share formula \eqref{eq:share} with $n = 10$, $\AUC = \tfrac12 + 3.0/10 = 0.80$ for A and $\tfrac12 + 2.2/10 = 0.72$ for B.
\begin{enumerate}[leftmargin=*]
\item \emph{AUC can prefer the wrong model.} Suppose a lender has funds for three loans, so it approves the three
applicants with the highest scores. Only $J_3$ matters: $0.2$ for A
and $0.6$ for B. AUC prefers A because A ranks well near the bottom of the list, where the lender makes no decisions.
\item \emph{Hand's point.} Under A, no lender would approve exactly three, since approving four adds a repayer and no
defaulter. Under B, approving three is what a lender with equal costs would do, since $J_3$ is B's largest value. Yet
AUC gives $J_3$ the same weight $1/10$ under both models. So the lenders that AUC averages over differ between the models.
\item \emph{$\mathrm{AUC}_\pi$.} Take $\pi = 1$ on the top three and $0$ elsewhere. Only pairs the policy splits, one
approved and one not, count. A split pair is ordered correctly when the repayer is the one approved.
\begin{itemize}
\item Model A approves $2$ repayers and $1$ defaulter, and leaves $3$ repayers and $4$ defaulters. It splits $11$ of
the $25$ pairs of a repayer and a defaulter: $2 \times 4 = 8$ correctly and $1 \times 3 = 3$ wrongly. So
$\mathrm{AUC}_\pi = 8/11 \approx 0.73$.
\item Model B approves $3$ repayers and no defaulters, and leaves $2$ repayers and $5$ defaulters. It splits
$3 \times 5 = 15$ pairs, all correctly. So $\mathrm{AUC}_\pi = 15/15 = 1$.
\end{itemize}
\end{enumerate}
\end{example}

The policy Youden index $J_\pi = \mathbb{E}[J(T_\pi)]$, the numerator of $\mathrm{AUC}_\pi$, is $J_3$ in the
example. It compares the models under
one fixed policy, so it is coherent. The denominator, the share of pairs the policy splits ($11/25$ for A, $15/25$ for
B), depends on the model, so the ratio is not.

The Brier skill score of the ideal forecast is also coherent, but its weights are not the decision maker's. It counts
every risk threshold equally, so a lender who approves only above a 90\% chance of repaying is scored mostly on
thresholds it never uses, and the Brier score also charges for miscalibration, which does not affect decisions based
on the ranking. $J_\pi$ puts the weight on the lender's own decisions and needs no forecast probability. The two agree,
up to scale, when the lender's probability of acting is linear in the true probability (Proposition~\ref{prop:link}).

In the example the ratio still orders the models correctly. Both models approve three applicants, so a model can do
better only by approving a repayer in place of a defaulter. That adds correct splits and removes wrong ones, so
$\mathrm{AUC}_\pi$ and $J_\pi$ rise together.

When the policy approves different numbers under different models, the ratio can reverse the order. Suppose the
lender approves everyone whose chance of repaying is above 70\%. Each model gives the applicants different chances of
repaying, so the number above 70\% can differ. Take the same ten applicants and two other models, C and D.
\begin{itemize}
\item Model C approves $1$ repayer and no defaulters, and leaves $4$ repayers and $5$ defaulters. So
$J_\pi = 1/5 - 0/5 = 0.2$. It splits $1 \times 5 = 5$ pairs, all correctly, so $\mathrm{AUC}_\pi = 5/5 = 1$.
\item Model D approves $3$ repayers and $1$ defaulter, and leaves $2$ repayers and $4$ defaulters. So
$J_\pi = 3/5 - 1/5 = 0.4$. It splits $14$ pairs: $3 \times 4 = 12$ correctly and $1 \times 2 = 2$ wrongly. So
$\mathrm{AUC}_\pi = 12/14 \approx 0.86$.
\end{itemize}
$J_\pi$ prefers D, which approves two more repayers at the cost of one more defaulter. $\mathrm{AUC}_\pi$ prefers C
only because C approves one applicant and so is judged on $5$ pairs, while D is judged on $14$. To compare models,
use $J_\pi$; $\mathrm{AUC}_\pi$ rescales it into a share of pairs.

The same problem arises when one model's AUC is compared across groups, such as two demographic groups in pretrial
risk assessment. Write $J_g$ and $\Fm_g$ for the Youden curve and the score distribution within group $g$, and
$S_{yg}$ for the score of a class-$y$ observation in group $g$. Within each group, Theorem~\ref{thm:auc} gives
$\AUC_g = \tfrac12 + \mathbb{E}[J_g(T_{\Fm_g})]$. The groups have different score distributions, so their AUCs weight
different cutoffs. A single policy, such as detaining everyone with a score above 7, is one cutoff for both groups,
and it falls at a different share of each group. Here the scores are on a common scale, so the fix is simple: choose
one weight $w$ for both groups and compare
\[
\mathbb{E}[J_g(T_w)] = \mathbb{E}[w(S_{1g})] - \mathbb{E}[w(S_{0g})],
\]
which by Lemma~\ref{lem:cutoff} needs only the mean of $w$ in each class of each group. The natural choice is the
policy itself, $w = \pi$, which compares the groups at the decisions actually made. Another possible choice is the
score distribution of the two groups combined, which keeps AUC's weighting by where the scores are but makes it the
same for both groups. This weight depends on the mix of the groups, and whether it is a sensible default remains to
be checked.

%% file: 08_conclusion/conclusion.tex
\section{Conclusion}
\label{sec:conclusion}

AUC, the Brier score and $\mathrm{AUC}_\pi$ are built on averages of the same curve, the Youden index $J(t)$, and differ in how
they weight the cutoffs. AUC weights them by where the scores are, the Brier score by the true probability, and
$\mathrm{AUC}_\pi$ by the decision maker's probability of acting. So when AUC and the Brier score rank two models
differently, part of the disagreement is about which decisions count.

This matters most when comparing models. AUC's weights come from each model's own scores, so two models' AUCs count
different decisions. The Brier score weights every model the same way, but its weights are not the decision maker's,
and it also charges for calibration. When the decision policy $\pi$ is known, models should be compared by the policy
Youden index $J_\pi$, the expected share of class 1 the policy acts on minus the expected share of class 0. It needs only the
mean of $\pi$ in each class. $\mathrm{AUC}_\pi$ puts this on the scale of a share of pairs, but its denominator depends on the
model, so it should not be used alone to rank models. When the policy is unknown, $J_\pi$ can be reported for a range
of plausible policies.

Selecting a model is this comparison on held-out data. Cross-validation keeps the candidate with the best score, so
the score decides which decisions count. If the score is AUC, the selected model can be one that ranks well where the
decision maker never acts. If it is $\mathrm{AUC}_\pi$, the denominator depends on the model, so it can reverse the order
$J_\pi$ gives (Section~\ref{sec:coherence}). When the policy is known, the score to maximize is $J_\pi$.
Hand's H-measure fixes a distribution of costs and scores each model at its own best cutoff. $J_\pi$ is the
criterion when the policy is already fixed and that cutoff will not be retuned.

Several questions remain. $J$ weights the two classes equally, which in cost terms means a cost ratio that depends on
the base rate \citep{smits2010note}. Extending the results to $\mathrm{TPR} - k\,\mathrm{FPR}$ is ongoing work. We do
not know whether another normalization of $\mathrm{AUC}_\pi$ keeps both coherence and the reading as a share of pairs.
Inference when $\pi$ is itself estimated remains to be worked out. The results also assume a binary outcome.

%% file: 09_appendix/appendix.tex
\section{Proofs}
\label{sec:proofs}

\phantomsection\label{prf:Jnonneg}
\begin{proof}[Proof of Lemma~\ref{lem:Jnonneg}]
By Bayes' rule,
\[
F_1(t) = P(S \le t \mid Y = 1) = F(t)\, \frac{P(Y = 1 \mid S \le t)}{P(Y = 1)}.
\]
Since $r$ is nondecreasing, $P(Y = 1 \mid S \le t) \le P(Y = 1 \mid S > t)$, and $P(Y = 1)$ is an average of the two, so
$P(Y = 1 \mid S \le t) \le P(Y = 1)$ and $F_1(t) \le F(t)$. The same argument for class 0 gives
$F_0(t) \ge F(t)$.
\end{proof}

\phantomsection\label{prf:weights}
\begin{proof}[Proof of Lemma~\ref{lem:weights}]
(a) Integrating by parts,
\begin{align*}
    \mathbb{E}[w(S_y)] &= \int w(s)\,dF_y(s) \\ 
     &=  \left[w(s)F_y(s)\right]^\infty_{-\infty} - \int F_y(t)\,w'(t)\,dt \\
     &= w(\infty) - \int F_y(t)\,w'(t)\,dt.
\end{align*}
Subtracting the two classes:
\[
    \mathbb{E}[w(S_1)]- \mathbb{E}[w(S_0)] = \int \left[F_0(t)-F_1(t)\right]\,w'(t)\,dt = \int J(t)\,w'(t)\, dt.
\]

(b) Since $S_y$ only takes the values $s_1, \dots, s_n$, only $w(s_1), \dots, w(s_n)$ enter $\mathbb{E}[w(S_y)]$. At those
points $w$ equals the staircase
\[
w(s_1) + \sum_{k=2}^{n} \left[w(s_k) - w(s_{k-1})\right]\ind\{s \ge s_k\}:
\]
at $s = s_j$ the sum telescopes to $w(s_j) - w(s_1)$. So we can treat $w$ as a step function that jumps by
$w(s_k) - w(s_{k-1})$ at each $s_k$, and
\begin{align*}
    \mathbb{E}[w(S_y)] &= \mathbb{E}\Bigl[w(s_1) + \sum_{k=2}^{n} \left[w(s_k) - w(s_{k-1})\right]\ind\{S_y \ge s_k\}\Bigr] \\
     &= w(s_1) + \sum_{k=2}^{n} \left[w(s_k) - w(s_{k-1})\right] P(S_y \ge s_k).
\end{align*}
Subtracting the two classes:
\begin{align*}
    \mathbb{E}[w(S_1)]- \mathbb{E}[w(S_0)] &= \sum_{k=2}^{n} \left[w(s_k) - w(s_{k-1})\right]\left[P(S_1 \ge s_k) - P(S_0 \ge s_k)\right] \\
    &= \sum_{k=2}^{n} \left[w(s_k) - w(s_{k-1})\right]\left[F_0(s_{k-1}) - F_1(s_{k-1})\right] \\
    &= \sum_{k=2}^{n} \left[w(s_k) - w(s_{k-1})\right] J(s_{k-1}).
\end{align*}
The second line uses that no score lies strictly between $s_{k-1}$ and $s_k$, so
$P(S_y \ge s_k) = P(S_y > s_{k-1}) = 1 - F_y(s_{k-1})$.
\end{proof}

\phantomsection\label{prf:cutoff}
\begin{proof}[Proof of Lemma~\ref{lem:cutoff}]
In the smooth case $T_w$ has density $w'$ on the real line and puts its remaining mass on $\pm\infty$, where $J = 0$,
so $\mathbb{E}[J(T_w)] = \int J(t)\, w'(t)\, dt$. In the discrete case $w$ is between-score, so $T_w$ puts no mass on the
scores and falls strictly between $s_{k-1}$ and $s_k$ with probability $w(s_k) - w(s_{k-1})$. Every cutoff there has
$J = J(s_{k-1})$. Its remaining mass lies below
$s_1$, above $s_n$ or at $\pm\infty$, where $J = 0$, so
$\mathbb{E}[J(T_w)] = \sum_{k=2}^{n} [w(s_k) - w(s_{k-1})]\, J(s_{k-1})$. In both cases this is
$\mathbb{E}[w(S_1)] - \mathbb{E}[w(S_0)]$ by Lemma~\ref{lem:weights}.
\end{proof}

\phantomsection\label{prf:auc}
\begin{proof}[Proof of Theorem~\ref{thm:auc}]
Let $w = (1 - \lambda)\Fm_0 + \lambda \Fm_1$ for some $\lambda \in [0,1]$. It is nondecreasing with values in $[0,1]$,
so with discrete scores its between-score version is a weight (Section~\ref{sec:tool}); with smooth class CDFs it is
already one. Taking $\lambda = \rho$ gives $w = \Fm$.

Let $S_0'$ and $S_1'$ be independent copies of $S_0$ and $S_1$, so that $\Fm_y(s) = P(S_y' < s) + \tfrac12 P(S_y' = s)$.
Evaluating at an independent score and taking expectations,
\begin{align*}
    \mathbb{E}[\Fm_0(S_1)] &= P(S_0 < S_1) + \tfrac12 P(S_0 = S_1) = \AUC, \\
    \mathbb{E}[\Fm_1(S_0)] &= P(S_1 < S_0) + \tfrac12 P(S_0 = S_1) = 1 - \AUC, \\
    \mathbb{E}[\Fm_0(S_0)] &= P(S_0' < S_0) + \tfrac12 P(S_0' = S_0) = \tfrac12, \\
    \mathbb{E}[\Fm_1(S_1)] &= P(S_1' < S_1) + \tfrac12 P(S_1' = S_1) = \tfrac12.
\end{align*}
The first line is the definition of AUC. The second holds because $\{S_1 < S_0\}$, $\{S_0 < S_1\}$ and $\{S_0 = S_1\}$
have total probability $1$. In the last two, $S_y'$ and $S_y$ have the same distribution, so $P(S_y' < S_y) = P(S_y < S_y')$,
and these two plus $P(S_y' = S_y)$ sum to $1$.
By linearity,
\begin{align*}
    \mathbb{E}[w(S_1)] &= (1 - \lambda)\AUC + \lambda\,\tfrac12, \\
    \mathbb{E}[w(S_0)] &= (1 - \lambda)\,\tfrac12 + \lambda\,(1 - \AUC).
\end{align*}
Subtracting the two classes:
\[
    \mathbb{E}[w(S_1)] - \mathbb{E}[w(S_0)] = (1 - \lambda)\bigl(\AUC - \tfrac12\bigr) + \lambda\bigl(\AUC - \tfrac12\bigr)
    = \AUC - \tfrac12.
\]
By Lemma~\ref{lem:cutoff} the left side is $\mathbb{E}[J(T_w)]$, so $\AUC = \tfrac12 + \mathbb{E}[J(T_w)]$ for every
$\lambda$, and $\lambda = \rho$ gives the theorem.
\end{proof}

\phantomsection\label{prf:brier}
\begin{proof}[Proof of Theorem~\ref{thm:brier}]
Write $r$ for $r(S)$. Since $Y$ is binary, $\mathbb{E}[Y] = \rho$ and $\mathrm{Var}(Y) = \rho(1-\rho)$.
\emph{$\mathrm{Var}(r) = \mathrm{Cov}(r, Y)$.} By the tower property, we first condition on $S$. Since $r$ is a function of $S$ with
$\mathbb{E}[Y \mid S] = r$,
\begin{align*}
    \mathbb{E}[rY] &= \mathbb{E}\bigl[\mathbb{E}[rY \mid S]\bigr] = \mathbb{E}\bigl[r\,\mathbb{E}[Y \mid S]\bigr] = \mathbb{E}[r^2], \\
    \mathbb{E}[Y] &= \mathbb{E}\bigl[\mathbb{E}[Y \mid S]\bigr] = \mathbb{E}[r].
\end{align*}
So
\[
    \mathrm{Cov}(r, Y) = \mathbb{E}[rY] - \mathbb{E}[r]\,\mathbb{E}[Y] = \mathbb{E}[r^2] - \mathbb{E}[r]^2 = \mathrm{Var}(r).
\]
Now we condition on $Y$ instead. Given $Y = y$, $S$ has the distribution of $S_y$, so
\begin{align*}
    \mathbb{E}[rY] &= \rho\,\mathbb{E}[rY \mid Y = 1] + (1 - \rho)\,\mathbb{E}[rY \mid Y = 0] = \rho\,\mathbb{E}[r(S_1)], \\
    \mathbb{E}[r] &= \rho\,\mathbb{E}[r \mid Y = 1] + (1 - \rho)\,\mathbb{E}[r \mid Y = 0] = \rho\,\mathbb{E}[r(S_1)] + (1 - \rho)\,\mathbb{E}[r(S_0)].
\end{align*}
In the first line, $rY = r$ when $Y = 1$ and $rY = 0$ when $Y = 0$. Then
\begin{align*}
    \mathrm{Cov}(r, Y) &= \mathbb{E}[rY] - \mathbb{E}[Y]\,\mathbb{E}[r] \\
    &= \rho\,\mathbb{E}[r(S_1)] - \rho\bigl[\rho\,\mathbb{E}[r(S_1)] + (1 - \rho)\,\mathbb{E}[r(S_0)]\bigr] \\
    &= \rho(1 - \rho)\,\mathbb{E}[r(S_1)] - \rho(1 - \rho)\,\mathbb{E}[r(S_0)] \\
    &= \mathrm{Var}(Y)\bigl(\mathbb{E}[r(S_1)] - \mathbb{E}[r(S_0)]\bigr).
\end{align*}
Under Assumption~\ref{ass:monotone} $r$ is a weight, so by Lemma~\ref{lem:cutoff} and the two steps
above,
\[
    \mathrm{Var}(r) = \mathrm{Var}(Y)\,\mathbb{E}[J(T_r)].
\]
Substituting into the Murphy decomposition,
\begin{align*}
    \BS &= \mathbb{E}\bigl[(p - r)^2\bigr] - \mathrm{Var}(r) + \mathrm{Var}(Y) \\
    &= \mathbb{E}\bigl[(p - r)^2\bigr] - \mathrm{Var}(Y)\,\mathbb{E}[J(T_r)] + \mathrm{Var}(Y) \\
    &= \mathbb{E}\bigl[(p - r)^2\bigr] + \mathrm{Var}(Y)\bigl(1 - \mathbb{E}[J(T_r)]\bigr). \qedhere
\end{align*}
\end{proof}

\begin{proof}[Proof of Corollary~\ref{cor:bss}]
For the ideal forecast $\BSS = \mathrm{Var}(r)/\mathrm{Var}(Y)$, which is $\mathbb{E}[J(T_r)]$ by the proof of Theorem~\ref{thm:brier}.
\end{proof}

\phantomsection\label{prf:aucw}
\begin{proof}[Proof of Theorem~\ref{thm:aucw}]
Since $w$ is nondecreasing:
\begin{itemize}[leftmargin=*]
\item if $S_0 < S_1$, then $\Delta_w \ge 0$, so $|\Delta_w|\,\ind\{S_0 < S_1\} = \Delta_w = \max(\Delta_w, 0)$;
\item if $S_0 > S_1$, then $\Delta_w \le 0$, so $|\Delta_w|\,\ind\{S_0 < S_1\} = 0 = \max(\Delta_w, 0)$;
\item if $S_0 = S_1$, then $\Delta_w = 0$, so $|\Delta_w|\,\ind\{S_0 < S_1\} = 0 = \max(\Delta_w, 0)$.
\end{itemize}
Therefore, $|\Delta_w|\,\ind\{S_0 < S_1\} = \max(\Delta_w, 0)$. Additionally, $\max(x, 0) = \tfrac12(x + |x|)$ for any real
$x$. So, by linearity,
\[
\AUCw = \frac{\mathbb{E}[\max(\Delta_w, 0)]}{\mathbb{E}|\Delta_w|}
= \frac{\tfrac12\,\mathbb{E}[\Delta_w] + \tfrac12\,\mathbb{E}|\Delta_w|}{\mathbb{E}|\Delta_w|}
= \frac12 + \frac{\mathbb{E}[\Delta_w]}{2\,\mathbb{E}|\Delta_w|}.
\]
Finally, by Lemma~\ref{lem:cutoff},
\[
\mathbb{E}[\Delta_w] = \mathbb{E}[w(S_1)] - \mathbb{E}[w(S_0)] = \mathbb{E}[J(T_w)]. \qedhere
\]
\end{proof}

\phantomsection\label{prf:link}
\begin{proof}[Proof of Proposition~\ref{prop:link}]
Let $v = \alpha + \beta w$ with $\beta > 0$. For any pair,
\begin{align*}
    \Delta_v &= v(S_1) - v(S_0) \\
    &= \bigl(\alpha + \beta w(S_1)\bigr) - \bigl(\alpha + \beta w(S_0)\bigr) \\
    &= \beta\bigl(w(S_1) - w(S_0)\bigr) \\
    &= \beta \Delta_w,
\end{align*}
and since $\beta > 0$, $|\Delta_v| = \beta |\Delta_w|$. In particular $\mathbb{E}|\Delta_v| > 0$ exactly when
$\mathbb{E}|\Delta_w| > 0$. Then, by linearity,
\[
\mathrm{AUC}_v
= \frac{\mathbb{E}\bigl[|\Delta_v|\, \ind\{S_0 < S_1\}\bigr]}{\mathbb{E}|\Delta_v|}
= \frac{\beta\,\mathbb{E}\bigl[|\Delta_w|\, \ind\{S_0 < S_1\}\bigr]}{\beta\,\mathbb{E}|\Delta_w|}
= \AUCw.
\]
Therefore $\AUCw$ is unchanged when $w$ is replaced by $\alpha + \beta w$.

Now let $\pi = \alpha + \beta r$. Taking $w = r$ above gives $\mathrm{AUC}_\pi = \mathrm{AUC}_r$. Under
Assumption~\ref{ass:monotone}, $r$ is a weight, so Theorem~\ref{thm:aucw} with $w = r$ and
$\Delta_r = r(S_1) - r(S_0)$ gives
\[
\mathrm{AUC}_\pi = \mathrm{AUC}_r = \frac12 + \frac{\mathbb{E}[J(T_r)]}{2\,\mathbb{E}|r(S_1) - r(S_0)|}.
\]
Finally, by Corollary~\ref{cor:bss}, $\mathbb{E}[J(T_r)]$ is the $\BSS$ of the ideal forecast, so the numerator is
that Brier skill score.
\end{proof}

\phantomsection\label{prf:pair}
\begin{proof}[Proof of Lemma~\ref{lem:pair}]
Check each ordering of the pair:
\begin{itemize}[leftmargin=*]
\item if $S_0 < S_1$: $\ind\{S_0 < S_1\} + \tfrac12 \ind\{S_0 = S_1\} = 1 = \tfrac12 + \tfrac12(1 - 0)$;
\item if $S_0 > S_1$: $\ind\{S_0 < S_1\} + \tfrac12 \ind\{S_0 = S_1\} = 0 = \tfrac12 + \tfrac12(0 - 1)$;
\item if $S_0 = S_1$: $\ind\{S_0 < S_1\} + \tfrac12 \ind\{S_0 = S_1\} = \tfrac12 = \tfrac12 + \tfrac12(0 - 0)$.
\end{itemize}
Therefore,
\[
\ind\{S_0 < S_1\} + \tfrac12 \ind\{S_0 = S_1\} = \tfrac12 + \tfrac12\bigl(\ind\{S_0 < S_1\} - \ind\{S_1 < S_0\}\bigr).
\]
Multiplying by $W(S_0, S_1)$, taking expectations and dividing by $\mathbb{E}[W(S_0, S_1)]$, by linearity,
\[
\mathrm{AUC}_W = \frac{\tfrac12\,\mathbb{E}[W(S_0, S_1)] + \tfrac12\,\mathbb{E}\bigl[W(S_0, S_1)\bigl(\ind\{S_0 < S_1\} - \ind\{S_1 < S_0\}\bigr)\bigr]}{\mathbb{E}[W(S_0, S_1)]},
\]
which is the formula.
\end{proof}

\phantomsection\label{prf:char}
\begin{proof}[Proof of Theorem~\ref{thm:char}]
For every distribution covered by Lemma~\ref{lem:weights}, Lemma~\ref{lem:cutoff} gives
$\mathbb{E}[J(T_w)] = \mathbb{E}[w(S_1)] - \mathbb{E}[w(S_0)]$. In the discrete case $w$ is replaced by its
between-score version (Section~\ref{sec:tool}), which has the same values at the scores.

\emph{If.} Let $W(s, s') = c\,|w(s) - w(s')|$ for all $s \ne s'$, and take any two numbers $s, s'$. Since $w$ is
nondecreasing:
\begin{itemize}[leftmargin=*]
\item if $s < s'$: $\ind\{s < s'\} - \ind\{s' < s\} = 1$ and $W(s, s') = c\bigl(w(s') - w(s)\bigr)$, so
$W(s, s')\bigl(\ind\{s < s'\} - \ind\{s' < s\}\bigr) = c\bigl(w(s') - w(s)\bigr)$;
\item if $s > s'$: $\ind\{s < s'\} - \ind\{s' < s\} = -1$ and $W(s, s') = c\bigl(w(s) - w(s')\bigr)$, so
$W(s, s')\bigl(\ind\{s < s'\} - \ind\{s' < s\}\bigr) = -c\bigl(w(s) - w(s')\bigr) = c\bigl(w(s') - w(s)\bigr)$;
\item if $s = s'$: $\ind\{s < s'\} - \ind\{s' < s\} = 0$ and $w(s') - w(s) = 0$.
\end{itemize}
So in every case $W(s, s')\bigl(\ind\{s < s'\} - \ind\{s' < s\}\bigr) = c\bigl(w(s') - w(s)\bigr)$. Putting $s = S_0$,
$s' = S_1$, taking expectations and using Lemma~\ref{lem:cutoff},
\[
\mathbb{E}\bigl[W(S_0, S_1)\bigl(\ind\{S_0 < S_1\} - \ind\{S_1 < S_0\}\bigr)\bigr]
= c\bigl(\mathbb{E}[w(S_1)] - \mathbb{E}[w(S_0)]\bigr) = c\,\mathbb{E}[J(T_w)]
\]
for every covered $F_0, F_1$.

\emph{Only if.} Let the identity hold for every covered $F_0, F_1$. Fix $s < s'$, and put all of class 0 at $s$ and all
of class 1 at $s'$; these distributions are discrete, so they are covered whether or not $w$ is smooth. Then
$S_0 = s < s' = S_1$ always, so
\begin{align*}
    \mathbb{E}\bigl[W(S_0, S_1)\bigl(\ind\{S_0 < S_1\} - \ind\{S_1 < S_0\}\bigr)\bigr] &= W(s, s'), \\
    c\,\mathbb{E}[J(T_w)] = c\bigl(\mathbb{E}[w(S_1)] - \mathbb{E}[w(S_0)]\bigr) &= c\bigl(w(s') - w(s)\bigr).
\end{align*}
The identity says these are equal, so $W(s, s') = c\bigl(w(s') - w(s)\bigr) = c\,|w(s) - w(s')|$, using that $w$ is
nondecreasing. Each pair $s < s'$ gets its own choice of distributions, so this holds for all $s < s'$. For $s > s'$,
$W$ is symmetric, so $W(s, s') = W(s', s) = c\,|w(s') - w(s)| = c\,|w(s) - w(s')|$.
\end{proof}

\phantomsection\label{prf:share}
\begin{proof}[Derivation of \eqref{eq:share}]
Let the scores $s_1 < \dots < s_n$ each have probability $1/n$. Then $P(S < s_k) = (k-1)/n$ and $P(S = s_k) = 1/n$, so
\[
\Fm(s_k) = \frac{k-1}{n} + \frac{1}{2n} = \frac{k - \tfrac12}{n},
\qquad
\Fm(s_k) - \Fm(s_{k-1}) = \frac1n.
\]
By Lemma~\ref{lem:cutoff} and Lemma~\ref{lem:weights}(b),
\[
\mathbb{E}[J(T_{\Fm})] = \sum_{k=2}^{n} \frac1n\, J(s_{k-1}).
\]
The cutoff $s_{k-1}$ acts on the scores above $s_{k-1}$, which are the top $n - k + 1$, so $J(s_{k-1}) = J_{n-k+1}$. As
$k$ runs from $2$ to $n$, $n - k + 1$ runs from $n - 1$ down to $1$, so
\[
\mathbb{E}[J(T_{\Fm})] = \frac1n\,(J_1 + \dots + J_{n-1}),
\]
and Theorem~\ref{thm:auc} gives \eqref{eq:share}.
\end{proof}